\documentclass[12pt]{iopart}

      \def\new#1 {{\bf #1 }}
      \def\cut#1 {\sout{#1} }

\def\HH {\hbox{${\rm H}_2$}}  %H2
\def\H2D {$\mathrm{H_2D^{+}}$} %H2D+
\def\simgreat{\mathbin{\lower 3pt\hbox
     {$\rlap{\raise 5pt\hbox{$\char'076$}}\mathchar"7218$}}}
\def\simless{\mathbin{\lower 3pt\hbox
     {$\rlap{\raise 5pt\hbox{$\char'074$}}\mathchar"7218$}}}

\def\aap {A\&A}
\def\apjl {ApJ. Letters}
\def\apj {ApJ.}
\def\apjs {ApJ.Suppl.}
\def\mnras {MNRAS}

\usepackage{natbib}
\usepackage{graphicx}
\bibpunct{(}{)}{,}{a}{}{;}
\usepackage{amssymb}
\usepackage{txfonts}
\begin{document}


\letter{183 GHz H$_2$O  maser emission around the low-mass protostar Serpens SMM1}


\author{T.A. van Kempen, D. Wilner, M. Gurwell}
\address{Center for Astrophysics,
    60 Garden Street, MS 78, Cambridge, MA 02138}
\ead{tvankempen@cfa.harvard.edu}

\begin{abstract}
We report the first interferomteric detection of 183 GHz water emission in the low-mass protostar Serpens SMM1 using the Submillimeter Array with a resolution of 3$''$ and rms of $\sim$7 Jy in a 3 km s$^{-1}$ bin. Due to the small size and high brightnessof more than 240 Jy/beam, it appears to be maser emission. In total three maser spots were detected out to $\sim$ 700 AU from the central protostar, lying along the red-shifted outflow axis, outside the circumstellar disk but within the envelope region as evidenced by the continuum measurements. Two of the maser spots  appear to be blue-shifted by about 1 to 2 km s$^{-1}$. No extended or compact thermal emission from a passively heated protostellar envelope was detected with a limit of 7 Jy (16 K), in agreement with recent modelling efforts.  We propose that the maser spots originate within the cavity walls due to the interaction of the outflow jet with the surrounding protostellar envelope. Hydrodynamical models predict that such regions can be dense and warm enough to invert  the 183 GHz water transition. 
\end{abstract}

\section{Introduction}

Water is one of the most important molecules in interstellar clouds in general and in star-forming regions in particular. In warm regions ($T>$100 K) close to the protostar, water is prominent with gas abundances up to 3$\times10^{-4}$ w.r.t. \HH, even higher than CO \citep{Cernicharo90,vanDishoeck96,Harwit98}. Besides the inner regions of protostellar envelopes these high abundances are also found where powerful jets from the protostar interact with the surroundings \citep{Nisini99}. In contrast, water abundances are as low as $10^{-8}$-$10^{-9}$ in cold ($T<$100 K) envelope regions as evidenced by Infrared Space Observatory (ISO) and Submillimeter Wave Astronomy Satellite (SWAS) observations \citep[e.g][]{Boonman03}.
Emission from  water molecules is very difficult to observe, due to the limits imposed by the Earth's atmosphere. 
Isotopologues such as deuterated water can be observed from the ground \citep[e.g.][]{Schulz91,Parise05} The only transitions of the main water isotope that can be observed regularly are low-frequency maser transitions. Most famous is the water maser at 22.2 GHz, the 6$_{16}$-5$_{23}$ transition, regularly observed using radio telescopes and interferometers, such as the Very Large Array (VLA). In star forming regions with embedded sources of high luminosity ($>$100 L$_{\odot}$) 22.2 GHz water maser emission is commonly detected and used to probe gas kinematics \citep[e.g.][]{Moscadelli06,Goddi06}. The emission is found to be variable on timescales of a day to a month, probably related to variations in the accretion disk or the outflowing material \citep[e.g.][]{Pashchenko05}. In regions with embedded sources of lower luminosity, 22.2 GHz water maser emission is detected less frequently than in high-mass sources, but again with considerable variability  \citep[e.g.][]{Wilking94,Claussen96,Claussen98,Furuya03}. Imaging with the VLA detected the  maser emission within several hundred AU of low-mass protostars \citep[e.g][]{Furuya99,Furuya01}, while Very Long Baseline Array observations indicate that the location may even be closer to the protostar \citep[e.g][]{Moscadelli06}.
 Both water and methanol masers have been observed to be related to disks \citep{Torrelles96,Moscadelli06}, while other observations of masers have been associated with outflows \citep[e.g.][]{Claussen96,Furuya03}.
 However, the excitation conditions for the 22 GHz water maser line are very high density ($>$10$^8$ cm$^{-3}$) and temperature (2000 K$>$ T $>$200 K)\citep{Yates97}, a combination of conditions that is rare in low-mass protostars. 

 Another water maser transition is found at 183.3 GHz. This 3$_{13}$-2$_{20}$ transition has been used in extra-galactic and galactic studies, primarily with the 30m telescope at Pico Veleta, Spain \citep[e.g.][]{Cernicharo94,Cernicharo96}. Statistical equilibrium calculations combined with the observation of extended emission conclude that relatively low temperatures (T$\sim$ 150 K) and densities (10$^5$-10$^6$ cm$^{-3}$) can already invert the populations of the 183.3 GHz transition \citep{Cernicharo94}.
In star forming regions it has been detected in several low- and high-mass protostellar sources such as Orion, W49N, H7-11 and L1448-mm. Most lines consist of a broad component,  superposed with a strong narrow line, presumably a maser line. The broad component was found to be spatially extended thermal emission after having been mapped in some high-mass star forming regions such as Orion and W49N \citep{Cernicharo94,GonzalesAlfonso95}. Extended thermal emission was also found  in low-mass sources such as HH7-11 \citep{Cernicharo96}. 
%Several studies have tried observing the 183 GHz maser with an interferometer, but these are mostly extra-galactic in nature \citep[e.g.][]{Humphreys05}.

Serpens SMM1 (referred to as SMM1) is a low-mass Class 0 source in the Serpens cluster ($D$= 250 pc) and has been studied at (sub)millimeter wavelengths by \citet{Hogerheijde99}. This source was selected as part of a pilot study to observe the 183 GHz line due to its inclusion in the Water in Star-forming regions with Herschel (WISH) program\footnote{See http://www.strw.leidenuniv.nl/WISH} as a source where many water lines will be targetted with spatial resolutions of 9-40$''$. It is relatively luminous ($L$=20.7 $L_\odot$) and thus an ideal target for water observations. It was observed with ISO-LWS \citep{Larsson02}, where numerous thermal water lines were found in the large 120$''$ beam. It drives a powerful highly collimated radio jet \citep{Curiel93}, a large molecular outflow \citep{White95} and was included in the 22 GHz water maser survey by \citet{Furuya03}, where several maser spots were found. These masers were in turn studied by \citet{Moscadelli06} who found the masers originating within a circumstellar disk based on their position and proper motions.  Observations with SWAS (beam = 3.3$'$ by 4.5$'$) of the 1$_{10}$-1$_{01}$ water line found water associated with the outflow with an abundance of o-H$_2$O of a few times 10$^{-7}$ \citep{Franklin08} in both red and blue outflow lobes.

In this letter, we present first interferometric  observations using the SMA\footnote{The Submillimeter Array is a joint project between the Smithsonian Astrophysical Observatory and the Academia Sinica Institute of Astronomy and Astrophysics and is funded by the Smithsonian Institution and the Academia Sinica.} of the 183 GHz maser emission from a region of low mass star formation, associated with the protostar SMM1. %The unique location of the SMA, on top of Mauna Kea, Hawaii, make it the only interferometer in the world to be able to carry out such a study.

%\section{Observations}

%Observations were taken in September 2008 using the SMA in compact configuration under excellent weather conditions (PWV $<$ 1 mm). The receiver was tuned to the water line at 183.310117 GHz. At this frequency and configuration combination, the beamsize is 3.1$''\times4.0''$. Due the opacity of the water in our atmosphere caused by pressure broadening, calibration purposes demanded  that the line was placed near the edge of the 2 GHz IF band to get the best signal at the other end of the IF band. As a first step the atmospheric water absorption across the IF band caused by the Earth's atmosphere was calibrated and removed by using highly smoothed observations on Jupiter. Due to a lower elevation than the actual source, the absorption was  overcorrected and additional calibration need to be performed on Uranus, Jupiter and nrao350, including small scale gain and bandpass variations. The calibration of the line itself was done using self-calibration on the brightest line in the spectrum of SMM1, as compared to the continuum flux level of 1751+096.  Due to the presence of water in our atmosphere, we estimate the flux calibration error to be 40$\%$, significantly larger than is normally the case. In total, the track consisted of only a single uninterrupted hour of usable data. The rest of the track was used to observe the 183 GHz water in N1333 IRAS 2A.

\section{Observations}

Observations were taken in September 2008 using the SMA in a compact
configuration(maximum baselines $\sim$75 m) under excellent weather conditions ( PWV $<$ 1 mm).  While the data set consisted of just one
uninterrupted hour of usable data for the target, SMM1, (RA=18:29:49.8, Dec=01:15:20.5) it was enough
to provide for unique spectral imaging of water masers.

The SMA frequency coverage includes separation of the two sidebands,
each covering 2 GHz with an IF coverage of 4-6 GHz. The receiver was
tuned to place the water line at 183.310117 GHz in the lower
sideband. At this frequency and configuration combination, the
beamsize is 3.1$''\times4.0''$. Due to pressure broadening, the
opacity of atmospheric water vapor is spread in a roughly gaussian
shape over several GHz, with a maximum at the rest frequency.  To
minimize the total average opacity across the full 2 GHz lower
sideband (and thus increase sensitivity to continuum emission near the
water line) the line was placed near one edge of the lower sideband,
providing the best possible sensitivity at the other end of the
sideband.

The passband shape is a mix of instrumental bandpass characteristics
and the atmospheric water absorption across the band. The instrumental
terms of the atmospheric absorption were calibrated and removed by using high
SNR observations of Jupiter. However, Jupiter was at a lower mean
elevation than SMM1, and thus the atmospheric absorption was
overcorrected for the target.  Additional calibration of the
atmospheric water vapor line shape was performed using spectrally
smoothed data from 1751+096, a nearby bright quasar.  Finally, under
the assumption that an individual maser component is spatially
unresolved, amplitude and phase self-calibration was performed using
the brightest maser feature from SMM1 to improve the imaging
reliability, though in the process we lose direct absolute position
information.  These absolute positions were retrieved by independently calibrating the upper sideband at 194 GHz using quasars and applying that solution to the already self-calibrated lower sideband.

As the observations were done near the center of the atmospheric water
line, we estimate the flux calibration error to be 40$\%$,
significantly larger than typical for observations well away from
water lines.  However, the errors in the relative fluxes are much smaller.

For continuum purposes, additional observations were taken in July 2009 using the SMA in very extended mode with baselines of up to $\sim$400 k$\lambda$. These observations were done at 220 GHz to avoid the water absorption and data were taken for 5 hours. Data were calibrated on 1751+096, with uranus and 3c454.4 as bandpass calibrator and Callisto as flux calibrator. 
 
\section{Results}
\subsection{Maser emission} 
The emission of the 183 GHz line shows a main line that is very bright (460 Jy/beam) with very narrow line-width (1 km s$^{-1}$) (Figure 1). Inspection of the map reveals two more line components, with brightnesses of  243 and 328 Jy/beam.  If one assumes that the emission fills the beam, the brightness temperatures are  1871,  982 and 1325 K respectively. Brightness temperatures are even much higher if one assumes a smaller source. Typical brightness temperatures of 22 GHz maser emission can be on the order of 10,000 K from very small regions. However, without higher resolution observations, it is impossible to conclude if these 183 GHz spots originate in similarly small spots or from larger regions. Even assuming it fills the beam, such temperatures are much higher than the ambient kinetic temperatures expected in protostellar envelopes and are thus not thermally excited.

\begin{figure}[!ht]
\begin{center}
\includegraphics[width=400pt]{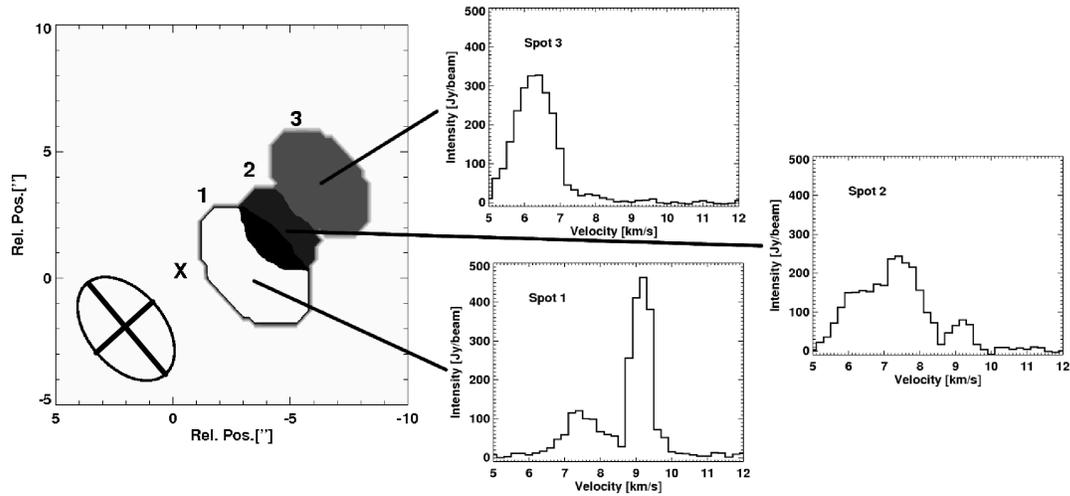}
\end{center}
\caption{Moment map of the three maser spots in SMM1. Maser spots are labeled with their respective number.  An X at the center 0,0 position marks the maximum intensity of the continuum (see Fig. \ref{1:cont}), which coincides with the location of the protostar. The spectra at the three maser spots are shown . Due to the proximity of the spot 2 to both spot 1 and 3 ($\sim$ 1 beamsize), the spectrum there is a blend of all three maser spots.  The beamsize is to 2.7$''$ by 4.0$''$ with a P.A. of 49$^\circ$.}
\label{1:moment}
\end{figure}

The maser spots are labelled '1', '2' and '3' in Figure \ref{1:moment}. Properties of the maser emission in each spot can be found in Table \ref{1:tab}, while spectra at each spot can be found in Fig. \ref{1:moment}. Several properties immediately jump out. First the center velocity of each of the spots decreases as a function of distance to the central protostar. Spot 3, which is the furthest away from the source, has the lowest velocity. Considering that the velocity of the Serpens SMM1 source itself is estimated at 8.7 km s$^{-1}$ from submillimeter single-dish and interferometric observations of molecular lines such as H$_2$CO and N$_2$H$^+$ \citep{Hogerheijde99}, this corresponds to the most blue-shifted velocity. Spot 1 is the only spot that is slightly red-shifted. However, the amount of both red- and blue-shifting is small. Second, the widths of the maser lines are not equal. The line associated with Spot 1 is a factor of 2 narrower than the other two, possibly indicating different excitation conditions.  Third, the spots seem to lie directly in the path of the red-shifted outflow as determined by \citet{Curiel93} and \citet{White95}. In \citet{Curiel93} it is also seen that at 3.6 cm continuum, small emission peaks appear at the red-shifted flow, but not the blue-shifted flow. These spots are at comparable positions as the maser emission (See also $\S$ 4)

\subsection{Thermal emission}
In the observations of Orion-KL, W49 and HH 7-11 \citep{Cernicharo94,Cernicharo96}, the 183.3 GHz emission also showed a broad emission profile that was thermal in nature. Mapping indeed showed that such thermal emission existed over larger areas. In our interferometric observations, we do not detect any thermal emission, down to a level of 7 Jy equal to 16.2 Kelvin, assuming it would fill the central beam. Due to the spatial filtering, it is possible that the emission is present at larger scales ($>$20$''$). However, this is not likely as in contrast with Orion, the interstellar radiation field near SMM1 is not strong with no more massive stars present. If one adopts the models from \citet{vanKempen08} and simulates the results through the beam and spatial filtering of the SMA, one expects a line of 6 to 20 K  within the central beam, equivalent to 2.5 to 8.4 Jy (depending on the combination of inner and outer water abundances, which range from 10$^{-4}$ to 10$^{-6}$ for material warmer than 100 K and 10$^{-6}$ to 10$^{-8}$ for material colder than 100 K). The emission predicted by this model is sufficiently compact, spanning less than 5$''$, that spatial filtering by the SMA would have no appreciable effect on a detection.

\subsection{Continuum}
\begin{figure}[!ht]
\begin{center}
\includegraphics[width=220pt]{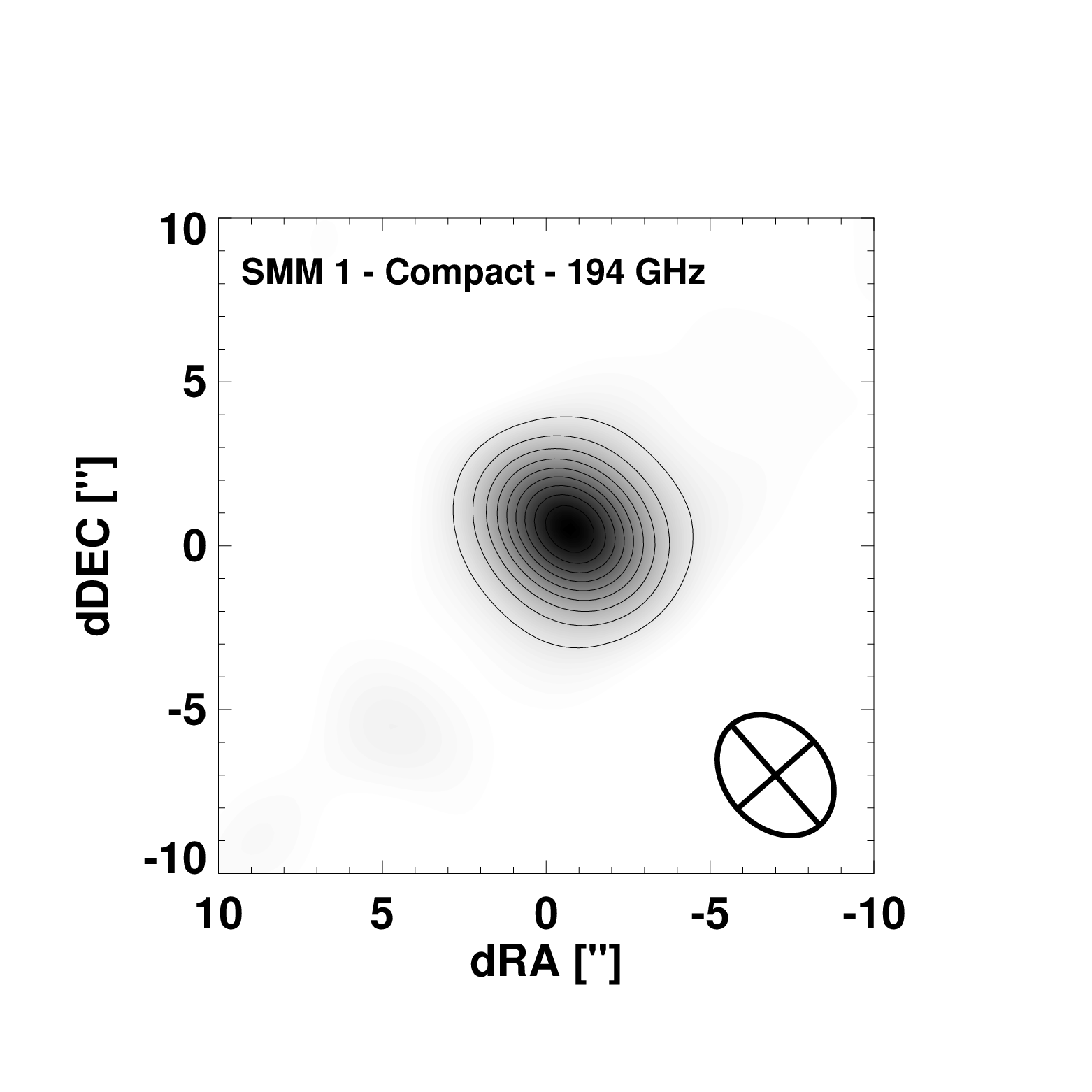}
\includegraphics[width=220pt]{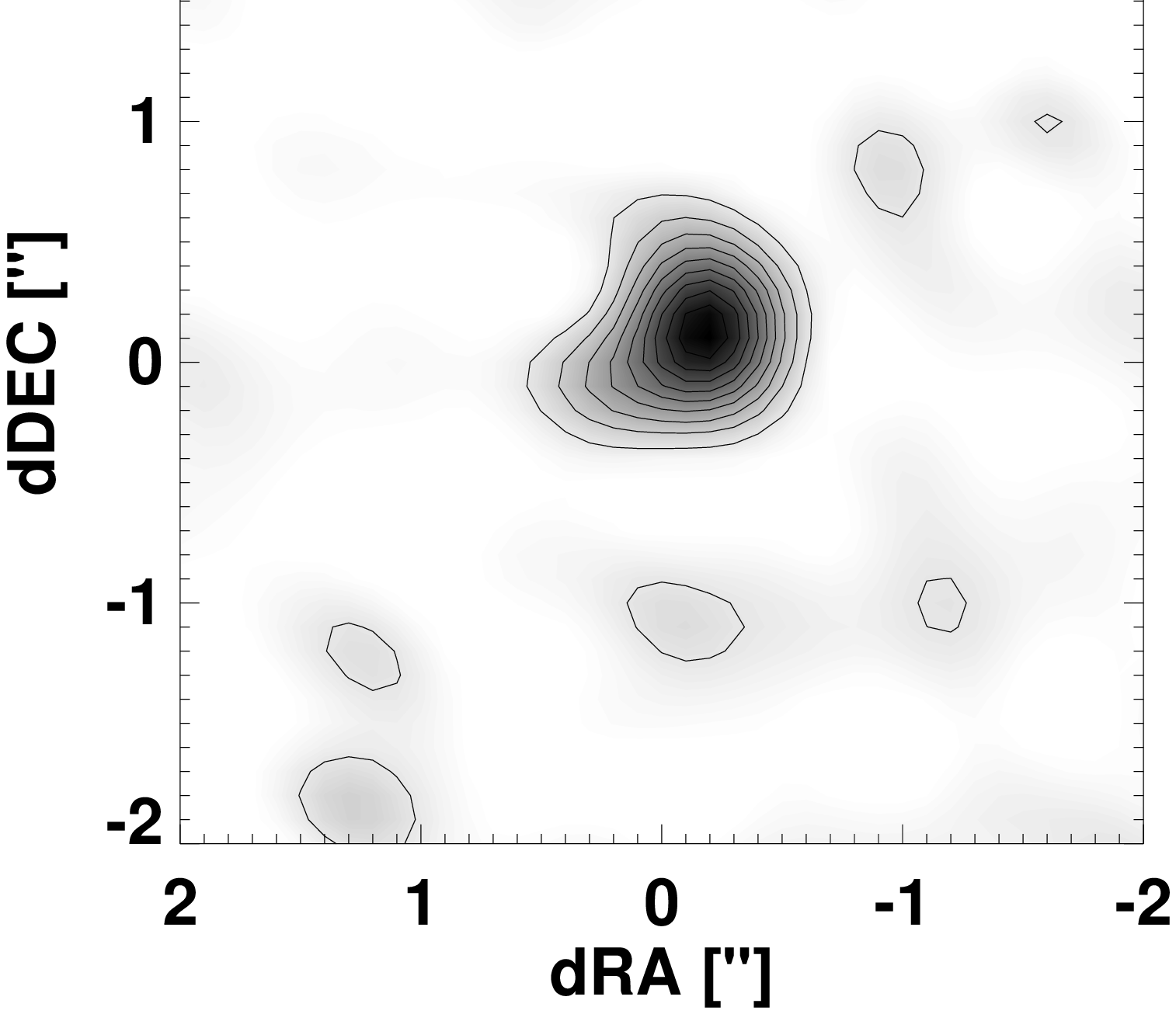}
\includegraphics[width=230pt]{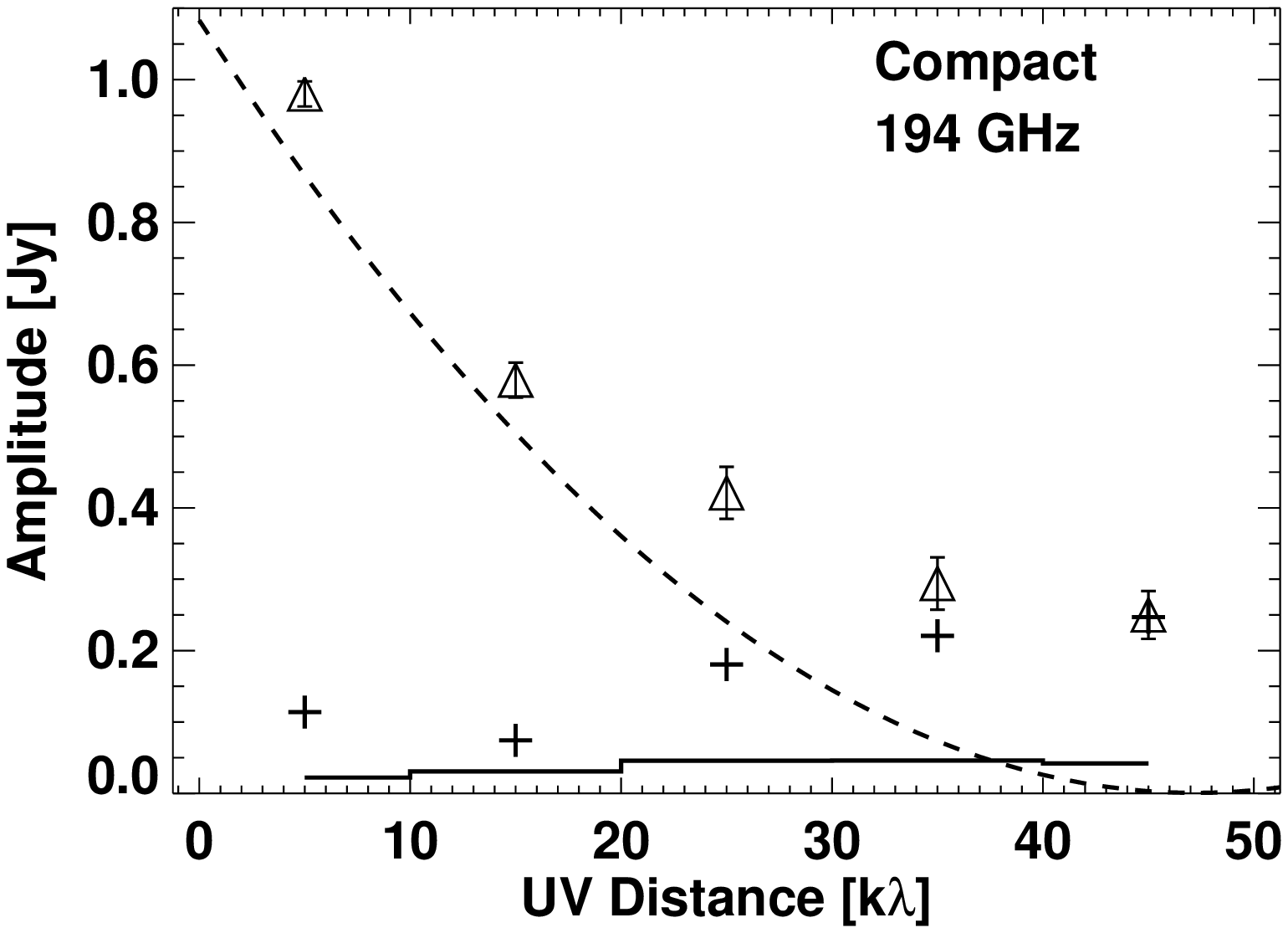}
\includegraphics[width=230pt]{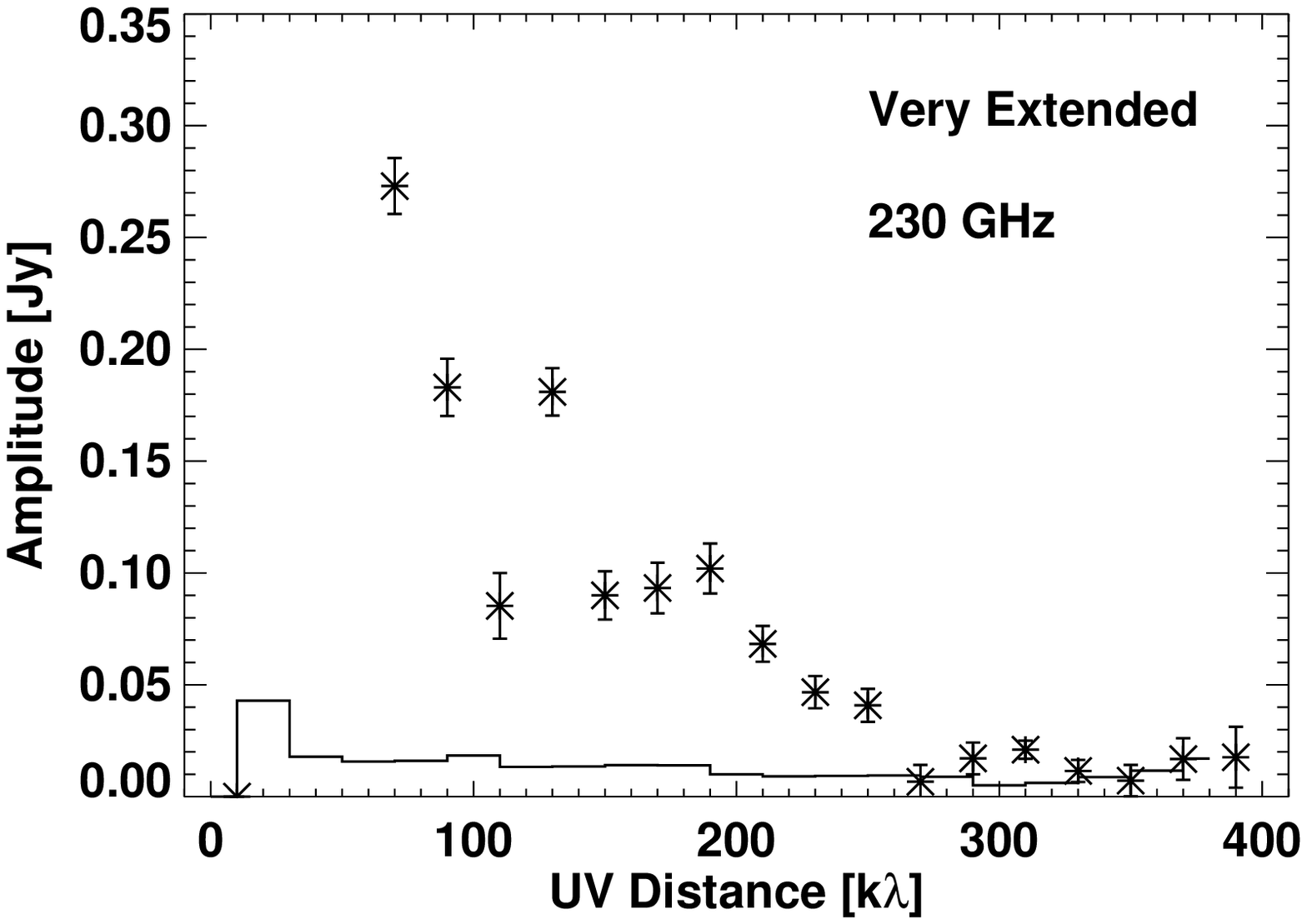}
\end{center}
\caption{{\it TOP:} The continuum map of SMM1 at 187 GHz.  Levels are in 2$\sigma$, 4$\sigma$, 6$\sigma$,... with $\sigma$=15 mJy/beam. {\it BOTTOM:} The UV distance versus amplitude plot, with the mean visibility amplitude in that annulus on the vertical axis and baseline length on the horizontal. The triangles show the observed visibilities of SMM1 at 187 GHz. Errors shown are the standard deviation in the mean, while the solid line shows the expectation value assuming no signal. The dashed line shows the expected visibilities for an envelope model at 194 GHz (Kristensen et al. in prep). The crosses represent the residual visibilities if the envelope model is subtracted from the observed visibilities.}
\label{1:cont}
\end{figure}

In addition to observing the line, the continuum was extracted from the same observations using the upper sideband at 194 GHz, which was free of lines. The resulting map is shown in the left part of  Fig. \ref{1:cont}, with the amplitude versus UV distance plot shown below it. The peak flux of the map is 0.52 Jy/beam. 
 Due to the limited UV-coverage, only 5 bins could be made. Overplotted is the expected visibility amplitude of a spherical model envelope based on SCUBA observations, modelled with Dusty \citep{Ivezic97} and sampled with the 194 GHz SMA coverage (Kristensen et al. in prep). This model has an envelope mass of 4.1 solar masses. The emission at longer baselines clearly is not fitted by the envelope model. Crosses represent the residual visibilities. If one ignores the drop in emission at the shortest wavelengths, this represents  a point source with a total emission of $\sim$ 0.55 Jy. An upper limit to the disk radius of $<$325 AU can deduced from this fit, which is in agreement with a compact source of $\sim$90 AU found by \citet{Hogerheijde99}. \\

Subsequently, observations were done of the continuum in a very extended configuration (k$\lambda$=400) with an rms of $\sim$2 mJy/beam. The resulting image is shown in the right side of the Fig. \ref{1:cont}. The disk is resolved there, with a radius of $\sim$100 AU. The plot of the amplitude versus UV distance shows that emission at baselines larger than 50 k$\lambda$ remains, until about 250 k$\lambda$ after which the sensitivity becomes too low for imaging the inner regions of the resolved disk. \citet{Moscadelli06} find the disk to be aligned in a NE-SW direction, but that is not confirmed .

The integrated flux of the compact emission in a 6$''$ radius is 2.3 Jy, with an associated mass of 2.7 M$_\odot$, assuming a temperature of 20 K and OH5 dust emissivities of \citet{Ossenkopf94}. The total dust mass of 4.1 M$_\odot$ also agrees well (Kristensen et al. in prep), as most of the cold dust emission on scales larger than 20$''$ is filtered out by the lack of baselines smaller than 10 k$\lambda$. \citet{Hogerheijde99} find a flux of 2.65 Jy at 1.4 mm (214 GHz), which is consistent with a $\nu^{3.5}$ dependence. Note that their mass of 8.7 M$_\odot$ is significantly higher due to the adopted distance of 400 pc.

If we assume that all the emission in the very extended image, which equals to 0.21 Jy, is originating in a resolved disk, then the total mass of the disk is $\sim$0.1 M$_\odot$, less than 1/20$^{\rm{th}}$ of the envelope mass. When compared to the sources observed in \citet{Jorgensen07}, one can conclude that although the envelope around SMM1 may be exceptionally massive compared to the disk and the disk on the small side, the observed profiles and masses are not unusual for a Class 0 source.
\section{Discussion}
\subsection{183 GHz maser excitation}
Two main studies have been carried out where the excitation conditions of the 183 GHz maser have been investigated.
Following \citet{Yates97}, the regime in which both 22 GHz and 183 GHz transitions can mase is at similarly high densities of 10$^{8}$ to 10$^{10}$ cm$^{-3}$ and temperatures of 200 to 2000 K. Such spots would be constrained to very small regions, probably about 1 AU in size at the maximum. However, a statistical equilibrium study of maser emission done by \citet{Cernicharo94} found different excitation conditions for the 183 GHz line, equivalent to densities of 10$^5$-10$^{6}$ cm$^{-3}$ and temperatures in excess of 50 K. Such conditions are much more prevalent near low-mass protostars, e.g. within the inner envelope regions, than those suggested by \citet{Yates97}.  For SMM1, the maser positions are a minimum of 675 AU away from the central protostar, a much larger distance than the size of the disk found in the UV-amplitude fitting (see above) and the location of the 22 GHz emission from \citet{Moscadelli06}, thus making it very unlikely that the two maser originate in the same region.
 The densities of the protostellar envelopes at the distances of the maser spots (+/- 700 to 1200 AU) are 10$^{6}$ cm$^{-3}$ \citep{Jorgensen02}. However, envelope temperatures derived by low-excitation submillimeter lines are generally much lower ($\sim$50 K), except in regions where the outflow and envelope interact.

\subsection{Maser origin}
From the emission profiles in Fig. \ref{1:moment}, the inherent properties of the 183.3 GHz maser excitation, comparison with the 22 GHz maser as well as the range of properties listed in Table \ref{1:tab}, the most likely origin for the emission of the 183 GHz masers is the cavity wall that the outflow jet creates by interacting with the spherical protostellar envelope. This can indeed be an origin for the 22 GHz maser line in high-mass protostars \citep{Goddi06}.  \citet{Delamarter00} and \citet{Cunningham05} performed detailed hydrodynamical simulations that show that outflows can sculpt the protostellar envelope, creating high-density regions ($>$10$^7$ cm$^{-3}$) in the cavity walls due to Kelvin-Helmholtz and Raleigh-Taylor instabilities. Temperatures at the cavity walls can reach a few hundred Kelvin from passing shocks and illumination by UV photons, an effect seen in recent observation of CO 6--5 and 7--6 \citep{vanKempen09}. Water will also be liberated from the grains, allowing for high abundances.
This environment, created uniquely in the cavity walls, is able to provide the necessary conditions for the populations to invert, creating the maser spots. A major unknown influence is probably the path length of the emission through the medium.

One problem however, is the velocity field, as the maser spots seem to be infalling instead of outflowing. As seen from \citet{Delamarter00} and \citet{Cunningham05}, the walls are very turbulent with infall and outflow material influencing each other in a Kelvin-Helmsholtz instability. It would pose little problem for some material in the cavity walls to be slightly infalling, explaining the presence of blue-shifted material in the red-shifted outflow. The changing path-lengths of the photons as they cavity walls evolve will also produce variability in the spots.

\section{Conclusion}
The conclusions of this letter are as follow
\begin{itemize}
\item The 183 GHz emission of Serpens SMM1 is masing from three distinct maser spots at 500-1200 AU, aligned in the direction of the red-shifted outflow, two of which are slightly blue-shifted.
\item The extended dust continuum emission detected with the SMA in Serpens SMM1 comes from a protostellar envelope with an estimated mass of 2.7 M$_\odot$.
\item A resolved component detected on baselines up to $\sim$400 k$\lambda$ indicates the presence of disk with an estimated mass of 0.1 M$_\odot$.
\item It is theorized that the maser spots originate in Kelvin-Helmholtz instabilities in the red-shifted outflow, created by the jet interactions with the envelope. Such interactions would temporarily create small regions of high density and temperature, providing the conditions for the 183 GHz transition to mase in low-mass protostellar environment.   
\item No compact or extended thermal water emission is detected, in agreement with a model of a passively heated envelope.
\end{itemize}
Research into the 183 GHz maser is promising as an additional constraint to jet mechanics and interaction of the jet with its surrounding material, both the envelope and the outflow.  In the next few years, the receivers of the SMA provide a unique opportunity to study the water maser emission at this wavelength in many low-mass YSOs.  In the long run, ALMA will be able to observe these lines at higher sensitivity and resolution, possibly probing the scales at which these masers emit. Combined with high-resolution VLA and VLBI observations of the 22 GHz maser, such observations can truly probe the physical structure of maser-emitting regions. 

\ack
TvK is supported as an SMA postdoctoral fellow. Steve Longmore and Jes J{\o}rgensen are thanked for discussion on fitting envelope models on continuum, and Elizabeth Humphreys for extensive discussion on maser excitation. TvK is also grateful to Lars Kristensen for providing the DUSTY model. The extensive ongoing discussions with Ewine van Dishoeck on many aspects of interstellar water are much appreciated.

\begin{table}[!ht]
\caption{Properties of the maser spots}
\begin{center}
\begin{tabular}{l l l l l l}
\hline \hline
Maser spot &  FWHM  & $I_{\rm{peak}}$  & $T_{\rm{B}}$$^a$ &$V_{\rm{LSR}}$ & Spat. Offset \\
& km s$^{-1}$ & Jy/beam & K &km s$^{-1}$ & RA,Dec ('','') \\ \hline
1 & 0.7 & 463& 1871 & 9.2 & -3.0,0.2  \\
2 & 1.5 & 243 & 982 &7.4 & -4.5,2.2\\
3 & 1.3 & 328 & 1325 & 6.3 & -6.7,3.8\\ \hline
\end{tabular}
\end{center}
$^a$ assuming the emission fills the beam
\label{1:tab}
\end{table}

\end{document}